\documentstyle[prl,aps,twocolumn,floats]{revtex}
 
\include{epsf}
\epsfverbosetrue

\def\e{{\, e}\,}

\def\be{\begin{equation}}
\def\ee{\end{equation}}
\def\bea{\begin{eqnarray}}
\def\eea{\end{eqnarray}}

\begin{document} 
%\newfloat{figure}{ht}{aux} 
%\draft
\twocolumn[\hsize\textwidth\columnwidth\hsize\csname
@twocolumnfalse\endcsname 

\title{Characterizing Phases of the non-Abelian 
Coulomb Gas}
\author{\bf Lori D. Paniak and Gordon W. Semenoff} 
\address{{\it Department of Physics and Astronomy, University of British
Columbia \\ 6224 Agricultural
Road, Vancouver, British Columbia V6T 1Z1}}  
\maketitle
 
\begin{abstract}
The thermodynamic 
problem of a gas of static quarks carrying U(N) charges
and interacting with each other via U(N) electric gauge fields is 
formulated and solved in the large N limit.  In a lattice theory, the solution can be found in any dimension.  In particular, in 1+1-dimensions, the continuum model can also be solved.  In that case, and when the quarks are in the adjoint representation, the explicit solution exhibits a first order quark confinement-deconfinement transition at a critical temperature and density.   
We also show that, when there are fundamental representation quarks, 
this phase transition persists until the relative density of fundamental 
quarks is comparable to the density of adjoint quarks, where it becomes 
a third order transition.   We discuss the possible interpretation of the 
third order transition as deconfinement. \\
{\it Presented by G.W.S at the 5th International Workshop
on Thermal Field Theories and their Applications, Regensburg, Germany, August
10-14, 1998.}  
\end{abstract} 

\vspace*{0.5truecm} ]

\section{Introduction}

One arena where non-perturbative QCD could be tested in the future is in
the study of nuclear matter in extreme conditions - high density or
high temperature.  An approximation to this situation will eventually
be produced in relativistic heavy ion collisions.  In contrast to the familiar confining phase of QCD which is relevant to nuclear physics and where 
the degrees of freedom are hadrons, which are the colorless bound states 
of quarks and gluons, at sufficiently
high temperature or density the system should be a plasma of the 
unconfined quarks and gluons themselves. Although there may not be a 
distinct phase transition
between these regimes, the existence of the quark gluon plasma should
have some well-defined signatures and studying its properties is
important.

At zero temperature, for states near the vacuum of QCD, it is very clear what 
confinement means - there are no asymptotic states of the quantum field theory
which carry color charge - all states are singlets of the global color symmetry. On the other hand, at finite temperature or density, where the physical 
system is in a mixture of states, there seems to be no clean 
characterization of confinement.  Part of the problem 
is the fact that the confining interaction is so strong at large distances that
it screens itself.  If a single test quark carrying color charge 
is introduced into a confining medium, there is a low energy state 
where the color charge of the test quark is screened by a dynamical 
anti-quark to make a color neutral object with finite energy.  

Of course, in a hypothetical system where quarks are absent, or where they 
occur only in the adjoint representation of the gauge group, so that they 
cannot screen the color charge of a fundamental representation 
test quark, one can characterize confinement by asking how
much energy it takes to introduce the test quark into the system.  In the confining phase it would be expected that it takes infinite energy, in the deconfined phase, finite energy.    

In the following, we will review a study of the issue of confinement
in a particular high temperature gauge theory which is a toy model of QCD. 
The idea is to study a model which is simple enough to be exactly solvable, 
but still complex enough to exhibit the phenomenon of interest - a phase transition between quark confining and a plasma phases.   The result 
of this study will be some speculations about characterizing the 
confining phase of a gauge theory at finite temperature and density when fundamental representation quarks are present.
The model with adjoint representation quarks can be
solved in any dimensions \cite{sz}.  However, here, we will review only
the one-dimensional case \cite{gps}.

In two dimensions, the Yang-Mills field itself has no propagating
degrees of freedom.  In adjoint QCD, the matter fields provide
dynamics by playing a role analogous to the transverse gluons of
higher dimensional gauge theory.  In fact, dimensional reduction of
three dimensional Yang Mills theory produces two dimensional QCD with
massless adjoint scalar quarks.  Moreover, since adjoint matter fields
do not decouple in the infinite $N$ limit, the large $N$ expansion is
of a similar level of complexity to that of higher dimensional
Yang-Mills theory.  One would expect it to exhibit some of the stringy
features of the confining phase which are emphasized in that limit.

Although adjoint QCD is not explicitly solvable, even at infinite $N$,
details of its spectrum were readily analyzed by approximate and
numerical techniques~\cite{dk,kutasov,Bhanot,kogan}.  In addition,
Kutasov ~\cite{kutasov} exploited an argument which was originally
due to Polchinski~\cite{polchinski} to show that the confining phase
must be unstable at high temperature and suggested it as a tractable
model where the confinement-deconfinement transition could be
investigated.

Here, we discuss a simplified version of 1+1 dimensional
QCD.\footnote{Generalization of these ideas to higher dimensions is
possible and a number of models are explicitly solvable there as well
\cite{sz}.  Here, for simplicity we concentrate on the one dimensional
case.} This model was formulated in ref.\cite{stz} and\cite{sz}.  We
consider a one-dimensional gas of non-dynamical particles which have
color charges and which interact with each other through non-Abelian
electric fields.

Even when the quarks are in the adjoint representation, this system
exhibits confinement.  Because there are no dynamical gluons which
could screen adjoint charges in one dimension, at low temperature and
density, adjoint quarks are confined\footnote{In higher dimensions, an
adjoint charge and a gluon could form a color singlet bound state.}.
They form colorless ``hadron'' bound states with two or more adjoint
quarks connected by non-dynamical strings of electric flux.  The large
$N$ limit resembles a non-interacting string theory in that the energy
of a state is proportional to the total length of all strings of
electric flux plus a chemical potential times the total number of
quarks.  The property of confinement is defined by estimating the
energy required to introduce an external fundamental representation
quark-antiquark pair into the system.  In the confining phase, where
the hadron gas is dilute, the quark-antiquark energy is proportional
to the length of the electric flux string which, to obtain gauge
invariance, must connect them.  This gives them a confining
interaction which increases linearly with their separation.

In the confined phase, the average particle number density and the
energy density are small --- in the large $N$ limit both are of order
one, rather than $N^2$ which one would expect from naive counting of
the degrees of freedom.  This is consistent with the fact that in a
confining phase the number of degrees of freedom, i.e.  hadrons, is
independent of $N$.  In contrast, in the deconfined phase, since the
number of degrees of freedom, i.e.  quarks and gluons, is proportional
to $N^2$ the particle density and energy are also of order $N^2$.

As temperature and density are increased, eventually we arrive at the
situation where inserting a quark-antiquark pair, involves a
negligibly small addition to the energy of the total system.  This is
the signature of the deconfined phase.

Between these two phases is a transition, which we show, is of first
order\cite{stz}.  In the string picture, this phase transition occurs
when the strings in a typical configuration percolate in the
one-dimensional space.  The order parameter is the Polyakov loop
operator ~\cite{pol,sus} which measures the exponential of the
negative of the free energy which is required to insert a single,
unpaired fundamental representation quark source into the system.
This free energy is infinite (and the expectation value of the
Polyakov loop is zero) in the confining phase and it is finite in the
deconfined phase.

The formalism we will use allows for a straightforward extension
of the model to include non-dynamical, fundamental representation
`quarks' \cite{gps} in addition to adjoint representation `gluons'.
QCD with fundamental representation quarks is solvable in the large
$N$ limit and has been used as an explicit example where
confinement-related phenomena can be studied\cite{thooft}.  Its
solvability follows partially from the fact that the quarks decouple
in the infinite $N$ limit.  We can also solve the non-Abelian coulomb
gas when it has fundamental representation quarks, which couple in a
non-trivial way when their fugacity is of order $N$. This resembles a
field theory when the number of flavors of heavy quarks, $N_F$, is of
the same order as the number of colors.  Then our solution applies to
the large $N$, $N_F\sim N$, heavy quark limit.

In a gas dominated by
fundamental representation particles, the energy needed to
introduce another fundamental representation quark is always finite,
and more sophisticated ideas are needed to study confinement
\cite{weiss}.  (For an interesting suggestion about symmetry breaking
in this case, see ref.\cite{gena}.)  Intuitively this is because the
electric flux associated with any external fundamental representation
source can be screened by a fundamental representation quark which is
already available in the system.  We will find that there is a third
order phase transition in this system, where the character of the
confining phase changes in a fundamental way.

The system we will consider here is that of a mixed gas of adjoint and 
fundamental representation non-Abelian charges interacting via 1+1
dimensional $U(N \rightarrow \infty)$ gauge fields.
The main result of our investigation is the development of a unified 
order parameter \cite{ldp} for discerning the low temperature confined phase 
from the high temperature quark-gluon plasma phase for all relative
densities of quarks and gluons.
This order parameter can be defined by considering the group theoretic
details of the mean-field solution of the model in the large $N$ limit.
We will outline this formalism and give some examples of its utility.

\section{Formalism}

The action of 1+1-dimensional Yang-Mills theory coupled to a number of
static particles is
\bea
S &=&
\int d^2x\left( -\frac{1}{e^2} {\rm Tr}F_{\mu\nu}F^{\mu\nu} \right. \\
&+& \left. i\sum_{j=1}^K
\ln{\rm Tr}{\cal P}\exp\left(i\int dt A_0(t,x_j)\right) \right)
\nonumber
\eea
In the canonical quantization of this system, the electric field is
the canonical conjugate of the spatial component of the gauge field,
\begin{equation}
\left[ A^a(x), E^b(y) \right]=i\delta^{ab}\delta(x-y)
\end{equation}
The Hamiltonian is\footnote{Here, for concreteness, we consider $U(N)$
gauge theory. The gauge field is $A = A^a t_a$, with $t_a$ the generators
in the fundamental representation. }
\begin{equation}
H=\int dx~ \frac{e^2}{2} \sum_{a=1}^{N^2} (E^a(x))^2 \ ,
\label{ham}
\end{equation}
and the temporal component of the gauge field, $A_0$ plays the role of
a Lagrange multiplier field which enforces the Gauss' law constraint,
\bea
\left( \frac{d}{dx}E^a(x) \right. &-& \left.
f^{abc}A^b(x)E^c(x) \right. \label{gauss} \\
&& \left. + \sum_{i=1}^K T^a_i\delta(x-x_i)\right)~\Psi_{\rm
phys.} =0 \nonumber
\eea
We will impose this constraint as a physical state condition.  The
particles with color charges are located at positions $x_1,\ldots
,x_K$.  $T^a_i$ are generators in the representation $R_i$ operating
on the color degrees of freedom of the {\it i}'th particle.

In the functional Schr\"odinger picture, the states are functionals of
the gauge field and the electric field is the functional derivative
operator $$ E^a(x)~=~\frac{1}{i}\frac{\delta}{\delta A^a(x)} $$ The
functional Schr\"odinger equation is that of a free particle
\bea
&& {\cal E}~\Psi^{a_1\ldots
a_K}\left[A;x_1, \ldots, x_K\right] \\
&=&  \int dx\left( -\frac{e^2}{2}\sum_{a=1}^{N^2}
\frac{\delta^2}{(\delta A^a(x))^2}\right)
~\Psi^{a_1\ldots a_K}\left[A;x_1, \ldots, x_K\right]
\nonumber
\eea
Gauss' law implies that the physical states, i.e. those which obey the
gauge constraint (\ref{gauss}), transform as
\bea
&& \Psi^{a_1\ldots a_K}\left[ A^g;x_1, \ldots, x_K\right] \\
&=&
g^{\rm R_1}_{a_1b_1}(x_1)\ldots g^{\rm R_K}_{a_Kb_K}(x_K)
\Psi^{b_1\ldots b_K}\left[A;x_1, \ldots, x_K\right] \nonumber
\eea
where $A^g\equiv gAg^{\dagger}+ig\nabla g^{\dagger}$ is the gauge
transform of $A$.  For a fixed number of external charges, this model
is explicitly solvable.  In the following we shall examine its
thermodynamic features, where we assume that the particles have
Maxwell-Boltzmann statistics.

We shall work with the grand canonical ensemble.  The partition
function is constructed by taking the trace of the Gibbs density
$e^{-H/T}$ over physical states.  This can be implemented by
considering eigenstates of $A^a(x)$ (and an appropriate basis for the
non-dynamical particles) $\vert A \rangle e_{a_1}\ldots e_{a_K}$.
Projection onto gauge invariant states involves a projection operator
which has the net effect of gauge transforming the state field at one
side of the trace, and integrating over all gauge transformations
\cite{gpy}. The resulting partition function is
\bea
&& Z[x_i,T] = \\
&& =\int[dA][dg]~\langle A \left| e^{-H/T} \right|A^g \rangle
{\rm Tr}~g^{\rm R_1}(x_1)\ldots
{\rm Tr}~g^{\rm R_K}(x_K) \nonumber
\eea
where $[dg(x)]$ is the Haar measure on the space of mappings from the
line to the group manifold and $[dA]$ is a measure on the convex
Euclidean space of gauge field configurations.  Here, we will consider
the case of particles in both the adjoint and fundamental
representations.  For U(N), the trace in the adjoint representation is
${\rm Tr}~g^{\rm Ad}(x) = \left| {\rm Tr}~ g(x)\right|^2$ where $g(x)$
is in the fundamental representation. In order to form the grand
canonical ensemble, we average over the particle positions by
integrating $\int dx_1\ldots\int dx_K$, multiply by the fugacity,
$\lambda$ for adjoint charges, $\kappa$ for fundamental charges,
$\kappa^*$ for the conjugate to the fundamental representation, to the
power of the number of respective charges, divide by the factorial
statistics factor to obtain Maxwell-Boltzmann statistics and sum over
all numbers of particles.  The result is
\begin{equation}
Z[\lambda,\kappa,T]~=~\int [dA][dg]~e^{-S_{\rm eff}[A,g]}
\end{equation}
where the effective action is
\bea
e^{-S_{\rm eff}[A,g]} 
&=& \langle A \left| e^{-H/T}\right| A^g
\rangle~ \\
& \times & \exp\left(\int dx~(\lambda\left|{\rm Tr}~ g\right|^2 +\kappa
{\rm Tr}~g+\kappa^*{\rm Tr}~g^{\dagger})\right)
\nonumber
\eea
The Hamiltonian is the Laplacian on the space of gauge fields.  Using
the explicit form of the  heat kernel
$$\langle A \left| e^{-H/T} \right| A^g\rangle \sim
\exp\left(-\int dx ~\frac{T}{e^2}~{\rm Tr}~(A-A^g)^2\right)\ ,$$
we see that the effective theory is the gauged principal chiral model
with a quadratic potential
\bea
&& ~~~~~~~~~~~~~~S_{\rm eff}[A,g] \label{partition}= \\
&& \int dx\left(\frac{N}{2 \gamma}~ {\rm Tr}\left|
\nabla g + i[A,g]\right|^2 -\lambda\left| {\rm
Tr}~g\right|^2  - 2 N \kappa ~{\rm Re} {\rm Tr} ~g \right) 
\nonumber
\eea
Here the integration over  gauge fields $A$ effectively 
enforces Gauss' law as one integrates over all elements of the 
gauge group with the Haar measure $[dg]$.
The fugacities of the adjoint and fundamental charges are
given by the parameters $\lambda$ and $N \kappa$, respectively.
Since we consider the matrix-valued fields $A$ and $g$ to 
be taken in the fundamental representation of $SU(N)$, 
the large $N$ limit will lead directly to the familiar 
situation of matrix models with large $N\times N$ matrices.
In order to keep all terms in the action of (\ref{partition})
at leading, $N^2$, order in this limit we will 
restrict parameters of the system such that
$\gamma \equiv  \frac{2T}{e^2 N} $, $\lambda$ and $\kappa$ 
are each of $O(1)$.

For the discussion of a confinement-deconfinement phase transition 
the most important aspect of the action in (\ref{partition})
is a global symmetry $S[A,g]~=$ $ S[A,~z~g]$
when the fundamental charge fugacity $\kappa$ vanishes. 
Here $z$ is a constant
element from the center of the gauge group, which for U(N) is U(1) and
for SU(N) is $Z_N$. It is this symmetry and its (thermo-)dynamical 
breaking that leads to the deconfinement phase transition in this model.
If $\kappa \neq 0$ the question of what remnants of this symmetry
persist is one we will answer in the next sections.
  
\begin{figure}
\epsfysize=2in
\epsfbox [60 245 530 535] {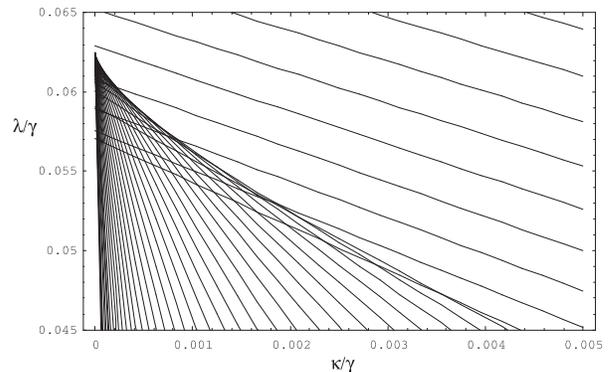}
\caption{ Plot of the lines (2.7) for $\mu$ ranging from 
$0.4$ (upper right corner) to $75$ (line at the extreme left).  
The region of overlapping lines corresponds to
a region of first order phase transition. \label{linesfig} }
\end{figure}

Additionally, there is a gauge invariance that can be used to diagonalize 
the matrices $g_{ij}(x)= e^{i\alpha_i(x)}\delta_{ij}$. 
The density of eigenvalues
$\rho(\theta,x)~=~\frac{1}{N}\sum_{i=1}^N \delta(\theta-\alpha_i(x)) $
corresponding to the large $N$ saddle-point evaluation of 
(\ref{partition}) now completely characterizes the
properties of the system.  Our goal is to find  this 
distribution  of eigenvalues.
Without loss of generality we can consider the Fourier expansion
\begin{equation}
\rho(\theta,x)~=~\frac{1}{2\pi}+\frac{1}{2\pi}
\sum_{n\neq 0}c_n(x) e^{-in\theta}
~,~c_n(x)^*= c_{-n}(x) 
\label{den}
\end{equation}
The configurations of the eigenvalue density
(\ref{den}) that saturate (\ref{partition}) at large $N$ can 
be found via the 
collective field theory approach \cite{JS,Zar}.   
The method is essentially based on the
relation between matrix quantum mechanics and non-relativistic
fermions \cite{bipz,Wadia}.  Leaving the details to \cite{gps}, it can be
shown that a solution of the saddle-point evaluation
of (\ref{partition}) is given by
\begin{equation}
\rho_0(\theta)=\left\{ \matrix{
\sqrt{\frac{8}{\gamma\pi^2} }\sqrt{ E+ 2 (\lambda c_{1}+\kappa)
\cos{ \theta} } &
\mbox{where } \rho \mbox{ is real} \cr 0 & {\rm otherwise}\cr}\right.
\label{Cdens}
\end{equation}
The constant of integration $E$ has physical interpretation as the
Fermi energy of a collection of $N$ fermions \cite{bipz}
on the circle subject to a periodic potential. It
is fixed by requiring the eigenvalue distribution to have 
unit normalization.
Furthermore, since the potential due to adjoint charges is 
non-local in eigenvalue space, 
the Fourier moment $c_1$ (see (\ref{den})) must be
self-consistently determined \cite{gubkleb}
\begin{equation}
c_1 \; = \;  \int d \theta ~\rho_0(\theta) \cos{\theta} 
\label{cnphi}
\end{equation}
This pair of conditions is most conveniently analyzed by
introducing a new parameter $\mu = E/(2(\lambda c_{1}+\kappa))$ and
the integrals over the positive support of $\mu +\cos{\theta}$,
\be 
I_n(\mu)=\frac{2}{\pi } \int d \theta \cos{n \theta} 
\sqrt{\mu +\cos{\theta}}
\ee
In terms of $\mu$, the solution of the normalization and moment
conditions is given by
\be 
c_1 = \frac{I_1(\mu)}{I_0(\mu)}
\ee
and 
\begin{equation}
\frac{{\kappa}}{\gamma} \; \; = \; \;  
\frac{1}{4 I_0(\mu)^2} - \frac{\lambda}{\gamma} \frac{ I_1(\mu)}{I_0(\mu)} 
\label{Cnecc}
\end{equation}
This last relation gives a family of lines in the $(\lambda /\gamma, 
\kappa/ \gamma)$ plane parameterized by $\mu$. As is shown in 
Figure \ref{linesfig} this family overlaps itself for lower densities
of fundamental charges, $\kappa / \gamma$ signaling the fact that
there are multiple solutions to the equations of motion in this region
of the phase diagram.
Considering the free energy one can show \cite{gps} that for lower
densities of adjoint charges the stable solution has $\mu>1$ while
at higher adjoint densities the stable solution has $\mu<1$.
In the intermediate regime there is a first order phase transition.
As the density of fundamental charges is increased the first order 
transition is smoothed out and a third order phase transition 
persists along the line $\mu=1$.

\begin{figure}
\epsfysize=3in
\epsfbox[50 241 490 648]{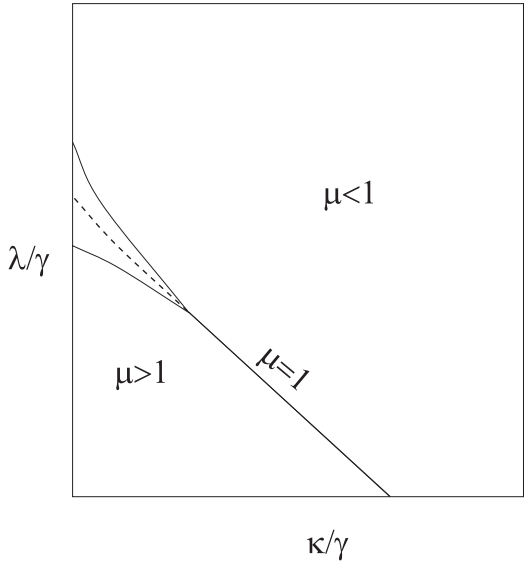}
\caption[Schematic picture of the phase diagram for the adjoint and 
fundamental representation non-Abelian Coulomb gas. ]
{Schematic picture of the phase diagram for the adjoint and 
fundamental representation non-Abelian Coulomb gas. 
The dotted curve marks the first order 
part of the critical line. The solid curves above and below it are the 
boundaries of the area with two possible phases. They join at a point
which shows second order behaviour. For larger $\kappa/\gamma$,
we find a third order line ($\mu = 1$) marked by a solid line.
\label{phasediag}} 
\end{figure}

The parameter $\mu$ is now seen to be useful for two different
reasons.  First it characterizes the general structure of the 
phase diagram (Figure \ref{phasediag}) where the `strong coupling' regime
is the region with $\mu>1$ and the `weak coupling' regime has
$\mu<1$.  As well, and of more importance for our analysis, 
we find that the expectation values of traces of powers of 
the group element $g$ are given as a function of the 
single parameter $\mu$
\be
\langle {\rm Tr}~g^n /N \rangle = c_n = \frac{I_n(\mu)}{I_0(\mu)}
\ee
Consequently, it makes sense for our purposes to re-define 
the eigenvalue distribution in terms of $\mu$
\be
\rho_0(\theta, \mu) = \frac{2}{\pi I_0(\mu)} 
\sqrt{ \mu + \cos{\theta}}
\label{rhomu}
\ee
In the next section we will use this definition and its connection 
to the dominant configuration of the gauge element to 
analyze the phase diagram in terms of group theory.

\section{Order parameters and the theory of group characters}
\setcounter{equation}{0}

As is known, in the case of pure gluo-dynamics, 
the realization of the center symmetry of the gauge group
governs confinement ~\cite{pol,sus}.  The Polyakov loop
operator ${\rm Tr}~ g(x)$, which is related to the free energy
$-T \log{ \langle\rm{Tr} g(x) \rm{Tr} g^\dagger (0)\rangle}$  of a 
conjugate pair of static, external fundamental charges  
separated by a distance $x$, serves as an order parameter 
\cite{sy} to test confinement. Since
${\rm Tr}~ g(x)$ transforms under the center as ${\rm
Tr}~g(x)~\rightarrow z~{\rm Tr}~g(x) $, the expectation value of
the Polyakov loop operator must average to zero if the 
center symmetry is preserved. Physically this suggests that
an infinite amount of energy is required to introduce
a single fundamental test charge into the system. The 
presence of a gas of fundamental charges   
$(\kappa \neq 0)$ changes this situation though by explicitly 
breaking the center symmetry. Consequently we lose the 
Polyakov loop operator as an order parameter for phase
transitions in the system. In this section 
we introduce a suitable
generalization of the Polyakov loop operator which will allow
us to identify a new order parameter.
 
As seen in the previous section, the
solution of the non-Abelian Coulomb gas with adjoint and 
fundamental representation charges is completely characterized 
by a Fourier sum of the 
traces $c_n= \langle{\rm Tr}~g^n/N\rangle$ - the higher winding Polyakov loops.
As noted in \cite{gps} the character of 
these traces changes between the strong and weak coupling regimes.
In particular, in the strong coupling $(\mu >1)$
phase $c_n$ is damped exponentially with $n$ while in the 
weak coupling $(\mu <1)$
phase the damping follows a power law behaviour.
Now we will reconsider this behaviour in terms of 
group theory. 

Since the matrix $g$ is an element of the special unitary group, 
its trace in an irreducible representation, $R$  
defines the group character for that representation
\be
\chi_R(g) \equiv {\rm Tr}_R~g
\ee
For the $N$ dimensional fundamental representation of $SU(N)$, $F$,
the group
character is just the Polyakov loop operator described above
since we are considering group elements to be taken 
in the lowest fundamental representation
\be
\chi_F(g) = {\rm Tr} ~g
\ee
Further simple examples are
the symmetric $(S)$ and anti-symmetric $(A)$ combinations
of a pair of fundamentals where we have
\be
\chi_S(g) =\frac{1}{2}[ ({\rm Tr}~ g)^2 + {\rm Tr} ~g^2 ]
\ee
and
\be
\chi_A(g) =\frac{1}{2}[ ({\rm Tr} ~g)^2 - {\rm Tr} ~g^2 ]
\ee
A general relation between characters and the group elements
is given by the Weyl formula but is not necessary for the 
following. A complete discussion can be found in 
standard references (see \cite{grpth} for example).
 
The main idea is that the eigenvalues of the group matrices, which 
are the only relevant dynamical variables in the grand partition 
function (\ref{partition}), are completely determined by the
$N$ quantities $\{ {\rm Tr} ~ g^n \}$, $n =1 \ldots N $.  In turn these traces 
form an algebraic basis equivalent to the characters of the 
$N$ fundamental (completely anti-symmetric)
irreducible representations of $SU(N)$ (including the 
trivial representation). 
Here we will explicitly
demonstrate the relationship between the basis of traces and the 
basis of group characters.
Ultimately it is the group theoretic variables which we will
use to characterize the phases of the model (\ref{partition}).
 
The standard basis for general
functions (of finite degree) of the eigenvalues of a matrix
is the set of elementary symmetric functions $\{a_r \}$.
In terms  of the eigenvalues  $\lambda_j =\e^{i \theta_j}$
of the group element $g$ they are given by
\bea
a_1 &=& \sum_j \lambda_j \\
a_2 &=& \sum_{j<k} \lambda_j \lambda_k  \nonumber \\
a_3 &=& \sum_{j<k<l} \lambda_j \lambda_k \lambda_l  \nonumber \\ 
\vdots  \nonumber \\
a_N &=&  \prod \lambda_j =\det{g} = 1
\eea
with $a_r \equiv 0$ for $r > N$.
The relationship of the symmetric functions $\{ a_r \}$ to the 
traces of the group elements, $S_n ={\rm Tr} ~ g^n$, is given \cite{lwd}
by the determinant 
\be
a_k = \frac{1}{k!} \left| \begin{array}{ccccccc}
S_1 & 1 & 0 &\cdots & & & \\
S_2 & S_1 & 2 & 0 &\cdots & & \\
S_3 & S_2 & S_1 & 3 & 0 &\cdots &  \\
\vdots & \vdots &\vdots &\vdots & & & 0 \\
S_{k-1}& S_{k-2} & S_{k-3} & \cdots & S_2& S_1& k-1 \\ 
S_k& S_{k-1}& S_{k-2} &\cdots & S_3& S_2& S_1   
\end{array} \right|
\label{bigdet}
\ee
Most importantly, it
can be shown  that the elementary symmetric functions are
nothing more than the characters of the fundamental representations
for the unitary group \cite{lwd,hagen}.  That is, for the 
fundamental representation which is the anti-symmetric combination
of $k$, $N$ dimensional representations, $\chi_k (g) = a_k$.

The determinant (\ref{bigdet}) can be evaluated \cite{muir}
in terms of a multinomial expansion most
compactly stated in terms of a generating function 
\be
\chi_{k}(g)=  \frac{(-1)^k}{k!} \frac{d^k}{dz^k}
\left. \exp{[ -\sum_{n=1}^{\infty} \frac{{\rm Tr} g^n}{n} z^n]}
\right|_{z=0}
\ee
For our purposes though it is useful to convert to a contour integral
about the origin. 
\be 
\chi_{k}(g) = \frac{(-1)^k}{2 \pi i } \oint \frac{dz}{z^{k+1}}
\exp{ \left[ -\sum_{n=1}^{\infty} \frac{{\rm Tr} g^n}{n} z^n \right]}
\label{genchar}
\ee
These last two expressions explicitly 
demonstrate the relationship between the 
group element $g$ and the $k^{th}$ fundamental representation of the 
gauge group and are completely general results.

With these relations we see that there is a direct connection
between the gauge group element $g$ and the irreducible 
(fundamental) representations of the gauge group.
In particular, in the previous section we have seen that
in the large $N$ solution of the non-Abelian Coulomb gas
a certain configuration of the gauge matrix, $g_0$ saturated 
the evaluation of the partition function (\ref{partition}).
Now it is natural to ask what is the configuration of 
irreducible representations corresponding to the dominant $g_0$.
This corresponds to evaluating the expectation
$\langle \chi_k(g)\rangle$ in the background of the non-Abelian gas.  
In principle this involves calculating expectations
of the form $\langle {\rm Tr} ~ g^{n_1} \cdots {\rm Tr}~ g^{n_r}\rangle$ but because of the 
factorization of gauge invariant objects in the limit $N \rightarrow 
\infty$, this reduces to a product of expectations, 
$\langle {\rm Tr} ~ g^{n_1}\rangle \cdots \langle{\rm Tr}~ g^{n_r}\rangle$.  Consequently 
$\langle \chi_k(g)\rangle$ is determined by replacing ${\rm Tr}~ g^n$
by its expectation value in (\ref{genchar}).  
Of course expectation values of the group element traces  are
intimately related to the eigenvalue density $\rho(\theta,\mu)$
(see (\ref{den})) hence, after performing an infinite sum, we obtain
\bea
&& \langle\chi_{\alpha}\rangle[ \rho(\theta, \mu) ] 
\equiv \frac{(-1)^{\alpha N}}{2 \pi i } \label{charint} \\
&& \times \oint \frac{dz}{z}
\exp{ \left[ \frac{N}{2} \int d \theta \rho(\theta,\mu) 
\log{\left(\frac{ 1 + z^2 - 2 z \cos{\theta}}{z^{2 \alpha}} \right)}
\right]}
\nonumber
\eea
Note that we have defined a new real parameter $\alpha= k/N$ on the 
unit interval that effectively 
labels the fundamental representations in the large $N$ limit.
Of course (\ref{charint}) now depends on a continuous variable
and is of a slightly different functional form than the discrete
case $\langle\chi_k\rangle$.  In the remainder of this discussion we will 
consider only the character parameterized by $\alpha$ as 
defined in (\ref{charint}).

\section{Calculation of the expectation of fundamentals}
\setcounter{equation}{0}
 
In this section we will concentrate on calculating 
$\langle\chi_{\alpha}\rangle$ 
with eigenvalue density (\ref{rhomu}) for the non-Abelian Coulomb gas.
This calculation will give a clear picture of the 
group theoretic excitations present in different regions of 
the phase diagram and consequently allow us to define 
an order parameter for the deconfinement transition even 
in the presence of fundamental matter.

Since explicit evaluation of (\ref{charint}) is difficult 
we begin with some special limiting cases.
As $\mu \rightarrow -1$ the support of the 
eigenvalue distribution (\ref{rhomu})
vanishes at $\theta =0$. The distribution does not vanish 
though as it retains
unit normalization and effectively becomes a delta function, 
$\delta(\theta)$.
Consequently we find the gauge matrix $g$ is just the identity
at $\mu=-1$, hence
\be
\langle \chi_\alpha \rangle = \lim_{N \rightarrow \infty}
\left( \begin{array}{c} N \\ \alpha N
\end{array} \right) =
2^N \sqrt{ \frac{2}{N \pi}} 
\e^{-2 N (\alpha -1/2)^2}
\ee
In this limit we find that the distribution of characters is 
symmetric about $\alpha =1/2$ as one would expect in a system 
where the total colour charge is vanishing.  As well in this limit
$\langle \chi_\alpha \rangle$ is non-vanishing and 
all fundamental representations are present in 
the large $N$ background solution of the model.  
As we will see, this result
is generic in the weak coupling phase $\mu <1$.

In the opposite limit, as $\mu \rightarrow
\infty$, it can be shown that the eigenvalue distribution 
(\ref{rhomu}) approaches a constant value $\rho=1/2 \pi$ with 
the eigenvalues of the group element $g$
becoming uniformly distributed on the unit circle.
Since expectation values of the traces of powers of 
the gauge matrix are essentially Fourier transforms of 
the eigenvalue distribution, it is easy to see that
$\langle{\rm Tr}~g^n\rangle \rightarrow 0$ in this limit and 
\be
\langle \chi_\alpha \rangle \rightarrow \delta_{0,\alpha }
\ee
This limit corresponds to the extreme strong coupling phase
of the model where the Polyakov loop operator $(\langle{\rm Tr}~g\rangle \sim 
\langle \chi_{1/N} \rangle )$ has vanishing 
expectation value and the standard analysis would point to 
a phase where colour charges are strictly confined into 
hadron-like structures.

In general the integral (\ref{charint}) can be evaluated by saddle-point
methods in the large $N$ limit in which we are interested.
The relevant action in this limit is
\be
S( \alpha, \mu, z)=\int d \theta \rho(\theta, \mu) 
\log{\left(\frac{ 1 + z^2 - 2 z \cos{\theta}}{z^{2 \alpha}} \right)}
\label{charS}
\ee
Solving the stationarity condition, $dS/dz |_{z_0} =0$,    
for $\alpha$ in terms of $z_0$ we find
the saddle-point condition for the large $N$ behaviour of the 
integral (\ref{charint}) is given by the relationship 
\be
\alpha = \int d \theta \rho(\theta, \mu) \frac{z _0 (z _ 0 - \cos{\theta})}
{1 + z_0^2 - 2 z_0 \cos{\theta}} 
\label{saddlept}
\ee
Since $\alpha$ is a real parameter restricted to 
the unit interval $[0,1]$ it can be shown 
that the saddle-point value of the parameter $z_0$ is real.
Further, for $z_0>1$  and $0<z_0<1$ Eqn.\ref{saddlept} returns values
of  $\alpha >1$ and  $\alpha <0$, respectively.  
Consequently we need only consider real, negative values of the 
parameter $z_0$. 

\begin{figure}
\epsfysize=2in
\epsfbox[65 327 398 535]{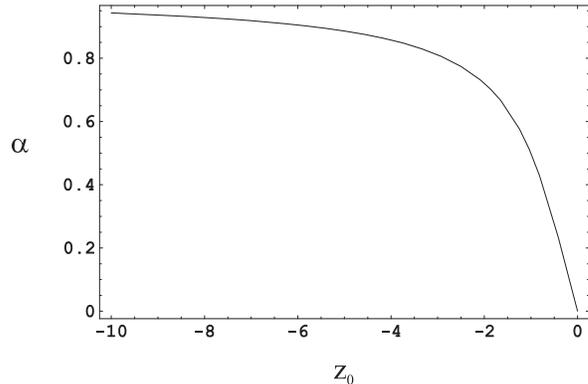}
\caption{Plot of the solutions of the saddle-point relation (7.60) for $\mu=0.5$.  \label{a05}}
\end{figure}

\begin{figure}
\epsfysize=2in
\epsfbox[120 329 480 533]{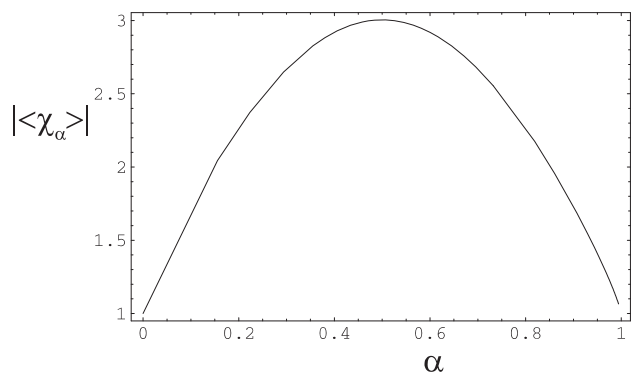}
\caption[Schematic diagram of $|\langle \chi_\alpha \rangle |$ vs. $\alpha$ for $\mu=0.5$.]
{Schematic diagram of $|\langle \chi_\alpha \rangle|$ vs. $\alpha$ for $\mu=0.5$. 
Note that all fundamental representations have non-vanishing expectation 
value.
\label{s05}}
\end{figure}

We now turn to an examination of the saddle-point approximation of
(\ref{charint})
for different regions of the phase diagram of the model at hand
beginning with the weak coupling phase, $\mu<1$. 
In this case the support of the 
eigenvalue distribution (\ref{rhomu}) is bounded away from 
$\theta =\pm \pi$ and hence the denominator in (\ref{saddlept})
is non-singular for all values of $z_0$. Consequently, in this regime
$\alpha$ varies smoothly and monotonically with $z_0$ and the relation (\ref{saddlept}) can in principle be inverted to obtain $z_0(\alpha)$.
With this information, the large $N$ 
asymptotic form of the expectation value of the 
characters $\langle \chi_{\alpha} \rangle$ can be determined by standard saddle-point 
methods.  In Figure \ref{a05} we show a numerically calculated example
of $\alpha$ as a function of $z_0$ for 
$\mu=0.5$. For this same case we
show a schematic diagram of the magnitude of the expectation value
$|\langle \chi_{\alpha} \rangle|$ as a function of $\alpha$ in Figure \ref{s05}.
In particular we see that the system has excitations in all irreducible 
representations.
 
\begin{figure}
\epsfysize=2in
\epsfbox[110 354 429 551]{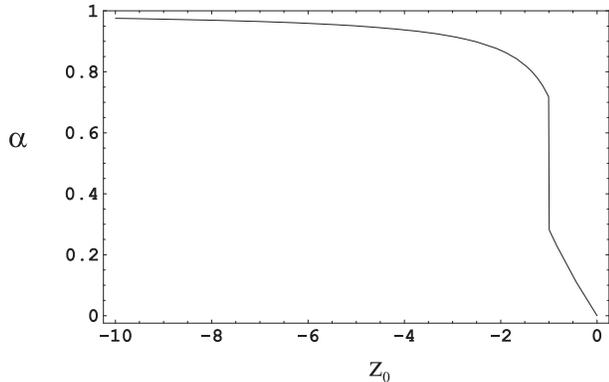}
\caption{Plot of the solutions of the saddle-point relation (7.60) for 
$\mu=1.2$.  \label{a12}}
\end{figure}

\begin{figure}
\epsfysize=2in
\epsfbox[-10 329 401 545]{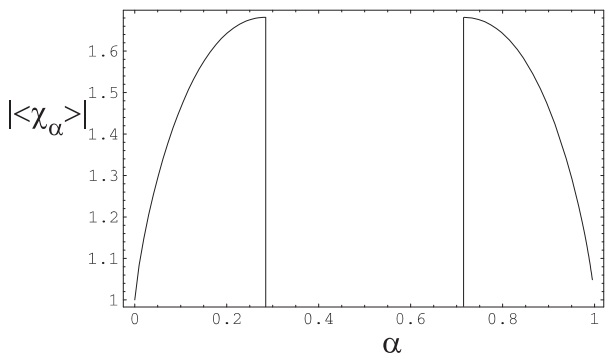}
\caption[Schematic diagram of $| \langle \chi_\alpha \rangle |$ vs. $\alpha$ for 
$\mu=1.2$]
{Schematic diagram of $|\langle \chi_\alpha \rangle|$ vs. $\alpha$ for 
$\mu=1.2$.  In this case the expectation value of representations with $\alpha$
between $\sim 0.25$ and $\sim 0.75$ is vanishing.  \label{s12}}
\end{figure}

For $\mu>1$ the situation is somewhat different. Now the 
support of the eigenvalue distribution (\ref{rhomu}) is 
the full interval $\theta \in [-\pi,\pi]$, and the denominator of 
(\ref{saddlept}) causes non-analytic behaviour to appear.
As one increases 
$\mu$ through unity the saddle-point relation for $\alpha$ shows this
non-analytic behaviour as 
a discontinuity at $z_0=-1$ (see Figure \ref{a12}). 
The result is that  
an open interval of $\alpha$ values centered on 
$\alpha =1/2$ are mapped into this discontinuity when the 
saddle-point relation (\ref{saddlept}) is inverted.  Since this 
discontinuity occurs in the saddle-point relation, it is 
not surprising to find that the curvature associated with 
the Gaussian integration of the saddle-point approximation
is divergent, effectively forcing the integral 
to vanish.  In terms of the expectation values of different
representations in the background of the non-Abelian Coulomb gas,
we see that an open interval of fundamental
representations centered about $\alpha =1/2$ is missing from 
the spectrum in the large $N$ limit.
In Figure \ref{s12} we show an example 
of the behaviour of the expectation value
$|\langle \chi_{\alpha} \rangle|$ with $\alpha$ for $\mu=1.2$.

The main outcome of this analysis is that the expectation value
of the central fundamental character $\langle\chi_{1/2}\rangle$ is vanishing 
if and only if $\mu \ge 1$. Consequently it may be considered
an order parameter distinguishing
between the strong and weak coupling phases of the model.  
Physically the situation is clear.  In the weak coupling phase
the system can effectively screen the interactions of any pair of
charges regardless of their representation since the system contains
excitations in all representations of the gauge group.
We conclude that the system 
looks much like a quark-gluon plasma where charges are effectively 
deconfined.  At the phase transition line
non-Abelian flux in the $\alpha=1/2$ fundamental representation 
becomes too energetically costly to produce and
the system can no longer screen the interaction between a pair
of $\alpha=1/2$ fundamental charges. In this strong coupling phase 
the interacting pair sees a linear confining potential (though
somewhat reduced as compared to the empty background).
As one further increases $\mu$ the
gap in the spectrum of fundamental representations becomes larger
and in the extreme limit $\mu = \infty$ the system contains only
excitations in the trivial representation.  This is precisely the 
confining phase of pure gluo-dynamics.
 
\section*{Acknowledgments}
This work was supported in part by the Natural Sciences and Engineering 
Research Council of Canada and a University of British Columbia 
Graduate Fellowship.

\end{document}